\documentclass{article}

\usepackage{cite}

\begin{document}

\date{}

\title{Tight upper and lower bounds to the eigenvalues and critical parameters of a
double well}

\author{Francisco M. Fern\'{a}ndez \thanks{E-mail: fernande@quimica.unlp.edu.ar} \\
INIFTA (CONICET, UNLP), Divisi\'on Qu\'imica Te\'orica\\
Blvd. 113 S/N, Sucursal 4, Casilla de Correo 16, 1900 La Plata,
Argentina}

\maketitle

\begin{abstract}
By means of a suitable rational approximation to the logarithmic derivative
of the wavefunction we obtain tight upper and lower bounds to the
eigenvalues and critical parameters of the quartic double-well potential.
\end{abstract}

\section{Introduction}

The anharmonic oscillator with the potential-energy function
$V(x)=m^{2}x^{2}+gx^{4}$ has been widely studied in many different
contexts and one can find hundreds of papers on this model that is
also discussed in many textbooks. It is therefore almost
impossible to try a satisfactory review of the literature. Here we
are mainly interested in some results obtained by
Turbiner\cite{T05,T10} who discussed both the single well
($m^{2}>0$, $g>0$) as well as the double well ($m^{2}<0$, $g>0$)
cases. In particular, Turbiner considered the most interesting
case in which $E(g_{crit})=0$ when $m^{2}=-1$. More precisely, in
the case of the double well the parameter $g$ is chosen so that
the energy equals the value of the potential at the top of the
barrier located between the two wells. Turbiner\cite{T05} obtained
$g_{crit}=0.302405$ for the ground state but it is obvious that
one can obtain similar critical potential parameters for every
quantum-mechanical energy. He also discussed the closely related
potential $V(x)=ax^{2}+2x^{2}$ for positive and negative values of
$a$\cite{T10}.

The purpose of this paper is to obtain several critical parameters
with sufficient accuracy using the Riccati-Pad\'{e} method (RPM)
that proved suitable for the calculation of accurate eigenvalues
of one-dimensional and separable Schr\"{o}dinger
equations\cite{FMT89a,FMT89b} (see also \cite{F08} for a
literature review). The RPM is based on a rational approximation
to the logarithmic derivative of the wavefunction. Although the
logarithmic derivative of the wavefunction is also the basis for
the remarkable variational-perturbation theory proposed by
Turbiner both approaches are completely different.

In section~\ref{sec:model} we briefly discuss the model,
section~\ref {sec:RPM} outlines the main ideas of the RPM in order
to make this paper sufficiently self-contained,
section~\ref{sec:results} contains a discussion of the results and
section~\ref{sec:comments} is devoted to further comments and
conclusions.

\section{The model}

\label{sec:model}

A symmetric double well $V(-x)=V(x)$ exhibits two minima of equal
depth $V(x_{m})$ at $x=\pm x_{m}$ and a barrier $V(0)>V(x_{m})$ at
$x=0$. The number of states with energies in the interval
$V(x_{m})<E\leq V(0)$ depends on the shape and depth $\left|
V(0)-V\left( x_{m}\right) \right| $ of the wells. For
concreteness, in what follows we consider the particular case
$V(x)=-x^{2}+gx^{4}$, $g>0$, for which $V(x_{m})=-1/(4g)$,
$x_{m}=1/\sqrt{2g}$ and $V(0)=0$. This potential also vanishes at
$x=\pm 1/\sqrt{g}$. The solutions to the Schr\"{o}dinger equation
\begin{equation}
\psi ^{\prime \prime }(x)+[E-V(x)]\psi (x)=0,  \label{eq:Schro}
\end{equation}
are either even $\psi (-x)=\psi (x)$ or odd $\psi (-x)=-\psi (x)$.
An even state exhibits a stationary point at $x=0$ because $\psi
^{\prime }(0)=0$. If we assume that $\psi (0)>0$ then the second
derivative at origin $\psi ^{\prime \prime }(0)=-E\psi (0)$
indicates that the point $x=0$ can be either a maximum, a minimum
or an inflexion point when $E>0$, $E<0$ or $E=0$, respectively.
Since $dE/dg=\left\langle x^{4}\right\rangle >0$ we conclude that
for every eigenvalue there exists a critical value $g=g_{crit}$
such that $E(g)>0$ when $g>g_{crit}$, $E(g)<0$ when $g<g_{crit}$
and $E(g_{crit})=0$. This analysis applies to every state $\psi
_{k}(x)$, where $k=0,1,\ldots $, and, consequently, we will have
the corresponding critical value $g_{k}$ given by
$E_{k}(g_{k})=0$, where $E_{0}<E_{1}<\ldots $ for every value of
$g$. However, in the case of and odd state we always have $\psi
(0)=0$ and $\psi ^{\prime \prime }(0)=0$ because of parity. As
indicated above, Turbiner\cite{T05,T10} calculated only the
critical value for the ground state.

\section{The Riccati-Pad\'{e} method}

\label{sec:RPM}

In this section we outline the application of the
RPM\cite{FMT89a,FMT89b} to the eigenvalue equation
(\ref{eq:Schro}) where $V(x)$ is symmetric about the origin:
$V(-x)=V(x)$. Without loss of generality we assume that $V(0)=0$.
If $V(x)$ is analytic at the origin we can expand it in a Taylor
series
\begin{equation}
V(x)=\sum_{j=1}^{\infty }V_{j}x^{2j}.  \label{eq:V_series}
\end{equation}
The regularized logarithmic derivative of the eigenfunction
\begin{equation}
f(x)=\frac{s}{x}-\frac{\psi ^{\prime }(x)}{\psi (x)},  \label{eq:f(x)}
\end{equation}
where $s=0$ or $s=1$ for an even or odd state, respectively, satisfies the
Riccati equation
\begin{equation}
f^{\prime }(x)+\frac{2sf(x)}{x}-f(x)^{2}+V(x)-E=0.  \label{eq:Riccati}
\end{equation}
Since the term $1/x$ in equation (\ref{eq:f(x)}) removes the pole
of $\psi ^{\prime }(x)/\psi (x)$ at the origin in the case of an
odd state, we can expand $f(x)$ in a Taylor series
\begin{equation}
f(x)=\sum_{j=0}^{\infty }f_{j}x^{2j+1},  \label{eq:f(x)_series}
\end{equation}
for any state (either even or odd). The coefficients $f_{j}$ can be easily
calculated by means of the recurrence relation
\begin{eqnarray}
f_{n} &=&\frac{1}{2n+2s+1}\left( \sum_{j=0}^{n-1}f_{j}f_{n-j-1}+E\delta
_{n0}-V_{n}\right) ,\;n=1,2,\ldots ,  \nonumber \\
f_{0} &=&\frac{E}{2s+1}.  \label{eq:fn}
\end{eqnarray}

The radius of convergence of the Taylor series (\ref{eq:f(x)_series}) is
determined by the zero of $\psi (x)$ closest to origin. A better
approximation is a rational function or Pad\'{e} approximant\cite{B75} that
takes into account all the zeroes of the eigenfunction (poles of $f(x)$).
However, instead of a standard Pad\'{e} approximant we choose the rational
approximation $x[M/N](x^{2})$, where
\begin{equation}
\lbrack M/N](z)=\frac{\sum_{j=0}^{M}a_{j}z^{j}}{\sum_{j=0}^{N}b_{j}z^{j}}%
=\sum_{j=0}^{M+N+1}f_{j}(E)z^{j}+O(z^{M+N+2}).  \label{eq:[M/N](z)}
\end{equation}
Since we can arbitrarily choose $b_{0}=1$ we are left with $M+N+1$
coefficients of the rational function and the unknown energy $E$
as independent adjustable parameters. Therefore, in order to
satisfy equation (\ref{eq:[M/N](z)}) the approximate energy should
be a root of
\begin{equation}
H_{D}^{d}(E)=\left|
\begin{array}{cccc}
f_{M-N+1} & f_{M-N+2} & \cdots  & f_{M+1} \\
f_{M-N+2} & f_{M-N+3} & \cdots  & f_{M+2} \\
\vdots  & \vdots  & \ddots  & \vdots  \\
f_{M+1} & f_{M+2} & \cdots  & f_{M+N+1}
\end{array}
\right| =0,  \label{eq:Hankel}
\end{equation}
where $d=M-N=0,1,\ldots $ and $D=N+1=2,3,\ldots $ is the dimension of the
Hankel determinant $H_{D}^{d}(E)$. The RPM is based on the fact that
sequences of roots $E^{[D,d]}$, $D=2,3,\ldots $ of the Hankel determinant
converge towards de actual eigenvalues of the Schr\"{o}dinger equation (\ref
{eq:Schro}).

In the case of the double-well potential $V(x)=-x^{2}+gx^{4}$ the Hankel
determinants depend on both $E$ and $g$. If we set $E=0$ then we obtain
sequences of roots $g^{[D,d]}$ that converge towards the actual critical
values of this potential parameter. It was shown that in the case of a
quartic potential there are sequences of roots of the Hankel determinants
that converge from above or below towards the eigenvalues depending on the
value of $d$\cite{FMT89a,FMT89b}. For this reason, when $E=0$ we expect to
obtain upper and lower bounds to the critical values of $g$.

\section{Results}

\label{sec:results}

In order to calculate the eigenvalues or their associated critical
parameters we obtain the coefficients $f_{j}(E,g)$ and the Hankel
determinants (\ref{eq:Hankel}) analytically by means of available
computer algebra software and then the roots of
$H_{D}^{d}(E=0,g)$. The numerical calculation of the roots is
straightforward because the coefficients $f_{j}(E,g)$ and,
consequently, the Hankel determinants are polynomial functions of
both $E$ and $g$.

\subsection{Critical parameters}

\label{subsec:critical}

Tables \ref{tab:g0}-\ref{tab:g11} show the critical parameters
$g_{k}^{[D,0]} $ and $g_{k}^{[D,1]}$ for $k=0,1,\ldots ,11$ and
$D\leq 30$. We appreciate that $g_{k}^{[D,0]}>g_{k}>$
$g_{k}^{[D,1]}$ which enables us to estimate the critical values
of $g$ with unprecedented accuracy because $\left|
g_{k}^{[D,0]}-g_{k}^{[D,1]}\right| \approx A_{k}e^{-\alpha
_{k}D}$, $\alpha _{k}>0$. Present results confirm the accurate
estimation of Turbiner\cite{T05}: $g_{0}=0.302405$. Although the
rate of convergence of the upper and lower bounds is exponential
and almost independent of $k$ the accuracy of the bounds for a
given $D$ decreases with $k$ because the sequence that approaches
$g_{k}$ starts at a determinant dimension $D_{k}$ that increases
with $k$. The reason is that the number of zeros of the
wavefunction increases with $k$ and, therefore, the degree of the
polynomial in the denominator of the rational approximation should
increase consistently to accommodate them.

We have decided to truncate all the results to $40$ digits even
when some of them have not converged. One can easily estimate the
actual accuracy of a particular result by straightforward
comparison of two consecutive terms of the sequence.

\subsection{Eigenvalues for a deep well}

\label{subsec:deep_well}

In the case of a deep well any approximation based on the Taylor
expansion of the wavefunction about the top of the barrier is
expected to be most inefficient\cite{T05,T10}. The RPM is not an
exception and its accuracy deteriorates noticeably with the well
depth. However, its rate of convergence is so great that it is
still a useful approach. In order to illustrate this point we
choose the potential $V(x)=ax^{2}+2x^{4}$ already studied by
Turbiner\cite{T10} who selected $a=-20$ as an illustrative example
of a deep well ($x_{m}=\sqrt{5}$, $V\left( x_{m}\right) =-50$).
Tables \ref {tab:e0} and \ref{tab:e1} show $E_{k}^{[D,0]}$ and
$E_{k}^{[D,1]}$ for $k=0,1 $, respectively, and $D\leq 30$. In
this case we appreciate that $E_{k}^{[D,0]}<E_{k}<E_{k}^{[D,1]}$
which enables us to estimate the eigenvalue with remarkable
accuracy because $\left| E_{k}^{[D,0]}-E_{k}^{[D,1]}\right|
\approx B_{k}e^{-\beta _{k}D}$, $\beta _{k}>0$. Once again we
confirm the accuracy of the results obtained by
Turbiner\cite{T10}: $E_{0}=-43.7793165$ and $E_{1}=-43.77931646$.

If we expand the potential about one of its minima, say $x=x_{m}$, then
there are two unknowns to be determined: $E$ and $f_{0}=f(x_{m})$. In this
case we have to resort to a variant of the RPM that is suitable for
nonsymmetric potentials\cite{FT96} that we do not discuss here in detail.
One of the features of the RPM is that the number of roots in the
neighbourhood of an eigenvalue increases with $D$\cite{F08}. This fact makes
the calculation of the optimal sequence of roots more difficult in the
present case because the search should be carried out in a two-dimensional
space. Table \ref{tab:e_ns} shows that the RPM does not yield neither $E_{0}$
nor $E_{1}$ but the average $\left( E_{0}+E_{1}\right) /2$ as is typical of
other approximations. An asterisk to the right of a result indicates that
the Newton-Raphson algorithm exhibited oscillatory behaviour and failed to
converge for that particular value of $D$ and $d$.

\subsection{Resonances}

\label{subsec:resonances}

For the model potential $V(x)=m^{2}x^{2}+gx^{4}$
Turbiner\cite{T05,T10} considered the cases $m^{2}>0$, $g>0$
(single well) and $m^{2}<0$, $g>0$ (double well). Another
interesting case is $m^{2}>0$, $g<0$ that supports complex
eigenvalues, or resonances, when the boundary conditions are those
for outgoing waves in both channels ($x\rightarrow -\infty $ and
$x\rightarrow \infty $). Since the RPM does not take the boundary
conditions explicitly into account the same Hankel quantization
condition (\ref {eq:Hankel}) also provides these complex
eigenvalues or resonances\cite{F95}. For example, when $m^{2}=1$
and $g=-0.1$ we obtain the accurate results shown in table
\ref{tab:e_res} for the lowest resonance. In this case the RPM
does not provide upper and lower bounds but the accuracy is also
remarkable.

\section{Further comments and conclusions}

\label{sec:comments}

The RPM has been applied to a wide variety of one-dimensional and separable
models along the years\cite{FMT89a,FMT89b,FT96,F95,F08}. In this paper we
show that it is suitable for the accurate calculation of the critical
parameters of parity-invariant double-well potentials. In the case of the
quartic potential it provides tight upper and lower bounds. The calculation
of the eigenvalues reveals another feature of the approach that was not
noticed earlier. If we apply the algorithm for symmetric potentials we
obtain tight upper and lower bounds to the eigenvalues even for the two
lowest ones that are almost degenerate in the case of deep wells. If, on the
other hand, we expand about one of the minima and apply the algorithm for
non-symmetric potentials\cite{FT96} we only obtain the average of such
energies.

In closing we want to stress once more the remarkable accuracy of the
results obtained by Turbiner\cite{T05,T10} by means of a
variational-perturbation approach based on relatively simple analytical
functions.

\begin{table}[H]
\caption{Upper and lower bounds to the critical parameter $g_0$}
\label{tab:g0}{\tiny {\ }}
\par
\begin{center}
{\tiny
\begin{tabular}{rll}
\hline
\multicolumn{1}{c}{$D$} & \multicolumn{1}{c}{$d=0$} & \multicolumn{1}{c}{$d=1
$} \\ \hline
2 & 0.3636964837266539687768291423593657530940 &
0.2916059217599021545472957231792727847579 \\
3 & 0.3041355199664415618743078042974272705046 &
0.3021387219255858306637347934008111792746 \\
4 & 0.3024440034802361046390034536436998252163 &
0.3023992995246179375766349606056642038148 \\
5 & 0.3024056418830358335891966561981441268327 &
0.3024047655333490990402062426326043560772 \\
6 & 0.3024048839946083646428716441326131478820 &
0.3024048682769592726748351355482080000580 \\
7 & 0.3024048703292784294366065525578006908341 &
0.3024048700650067865802797647536576226344 \\
8 & 0.3024048700986220241695241682236655194508 &
0.3024048700943923832247301685376725595064 \\
9 & 0.3024048700949194356723287031392559699348 &
0.3024048700948543313895839076211758857242 \\
10 & 0.3024048700948623100521380286778238786326 &
0.3024048700948613392649500863709288207287 \\
11 & 0.3024048700948614566082862233924157760399 &
0.3024048700948614425100082233465182363622 \\
12 & 0.3024048700948614441944313418759149633306 &
0.3024048700948614439942175162248892570706 \\
13 & 0.3024048700948614440179014941243477196370 &
0.3024048700948614440151123201520973863135 \\
14 & 0.3024048700948614440154394233185815060767 &
0.3024048700948614440154012115553250387643 \\
15 & 0.3024048700948614440154056590605573924369 &
0.3024048700948614440154051431943447821851 \\
16 & 0.3024048700948614440154052028352994883999 &
0.3024048700948614440154051959611789569142 \\
17 & 0.3024048700948614440154051967511775059455 &
0.3024048700948614440154051966606380877527 \\
18 & 0.3024048700948614440154051966709874069051 &
0.3024048700948614440154051966698073464149 \\
19 & 0.3024048700948614440154051966699415817037 &
0.3024048700948614440154051966699263465505 \\
20 & 0.3024048700948614440154051966699280719473 &
0.3024048700948614440154051966699278769474 \\
21 & 0.3024048700948614440154051966699278989422 &
0.3024048700948614440154051966699278964660 \\
22 & 0.3024048700948614440154051966699278967443 &
0.3024048700948614440154051966699278967131 \\
23 & 0.3024048700948614440154051966699278967166 &
0.3024048700948614440154051966699278967162 \\
24 & 0.3024048700948614440154051966699278967162 &
0.3024048700948614440154051966699278967162 \\
25 & 0.3024048700948614440154051966699278967162 &  \\ \hline
\end{tabular}
}
\end{center}
\par
{\tiny \ }
\end{table}

\begin{table}[H]
\caption{Upper and lower bounds to the critical parameter $g_2$}
\label{tab:g2}{\tiny {\ }}
\par
\begin{center}
{\tiny
\begin{tabular}{rll}
\hline
\multicolumn{1}{c}{$D$} & \multicolumn{1}{c}{$d=0$} & \multicolumn{1}{c}{$d=1
$} \\ \hline
4 &  & 0.07249401976717864151512732239458411564730 \\
5 & 0.07880383087757011137683529737824261495507 &
0.07753054612942325838337430417287006369320 \\
6 & 0.07777666090366403931216840109390188166990 &
0.07773099784924783071543382424333385673545 \\
7 & 0.07773920540992483611759533631182864744840 &
0.07773777294782089944699396518698855106312 \\
8 & 0.07773801660196284966767006960062932769463 &
0.07773797610483737979072405485459871898530 \\
9 & 0.07773798269725003135521198565186191728902 &
0.07773798164410960093923147751655311453369 \\
10 & 0.07773798180949031979011418493561074967290 &
0.07773798178392349671164663015473467484454 \\
11 & 0.07773798178781948377703152449820346966539 &
0.07773798178723362245540683204039908883939 \\
12 & 0.07773798178732064652096414095882960096355 &
0.07773798178730786644740643392685396952654 \\
13 & 0.07773798178730972347206342063332123749390 &
0.07773798178730945629590258954930069113686 \\
14 & 0.07773798178730949438039534093379878518226 &
0.07773798178730948899872407316304388880808 \\
15 & 0.07773798178730948975299872864487236421486 &
0.07773798178730948964809530047846122383613 \\
16 & 0.07773798178730948966257915559793548142474 &
0.07773798178730948966059313555714627470520 \\
17 & 0.07773798178730948966086368372390240697374 &
0.07773798178730948966082705590320204467144 \\
18 & 0.07773798178730948966083198555066043390927 &
0.07773798178730948966083132579721610049862 \\
19 & 0.07773798178730948966083141362302671731686 &
0.07773798178730948966083140199133518119823 \\
20 & 0.07773798178730948966083140352433436503540 &
0.07773798178730948966083140332323498065259 \\
21 & 0.07773798178730948966083140334949739651516 &
0.07773798178730948966083140334608237440235 \\
22 & 0.07773798178730948966083140334652461846449 &
0.07773798178730948966083140334646757442932 \\
23 & 0.07773798178730948966083140334647490443053 &
0.07773798178730948966083140334647396599301 \\
24 & 0.07773798178730948966083140334647408571461 &
0.07773798178730948966083140334647407049300 \\
25 & 0.07773798178730948966083140334647407242196 &
0.07773798178730948966083140334647407217829 \\
26 & 0.07773798178730948966083140334647407220897 &
0.07773798178730948966083140334647407220512 \\
27 & 0.07773798178730948966083140334647407220560 &
0.07773798178730948966083140334647407220554 \\
28 & 0.07773798178730948966083140334647407220555 &
0.07773798178730948966083140334647407220555 \\
29 & 0.07773798178730948966083140334647407220555 &
0.07773798178730948966083140334647407220555 \\ \hline
\end{tabular}
}
\end{center}
\par
{\tiny \ }
\end{table}

\begin{table}[H]
\caption{Upper and lower bounds to the critical parameter $g_4$}
\label{tab:g4}{\tiny {\ }}
\par
\begin{center}
{\tiny
\begin{tabular}{rll}
\hline
\multicolumn{1}{c}{$D$} & \multicolumn{1}{c}{$d=0$} & \multicolumn{1}{c}{$d=1
$} \\ \hline
7 & 0.04637676319118771829167960772482706323306 &
0.04448266941691055535466619198018892670832 \\
8 & 0.04486554986896052877360988330875349112100 &
0.04478721704291366918609339348372248917650 \\
9 & 0.04480270069395284067189801456884715258892 &
0.04479971395760984967156162050942791454765 \\
10 & 0.04480027677096987196489514247594614354609 &
0.04480017295777737728036026969215278345688 \\
11 & 0.04480019173271378075989120345689529452345 &
0.04480018839860184546193232659737398313455 \\
12 & 0.04480018898073987718712336666213537971711 &
0.04480018888068847861919524987787867883116 \\
13 & 0.04480018889763290676251510452250331007217 &
0.04480018889480251171492689131961124002445 \\
14 & 0.04480018889526922906221996330640653139816 &
0.04480018889519319933956676058573340904971 \\
15 & 0.04480018889520544383601354410532360734255 &
0.04480018889520349306976270269563966644968 \\
16 & 0.04480018889520380070164273753041430874870 &
0.04480018889520375265608819979371438996645 \\
17 & 0.04480018889520376009117480127738216852773 &
0.04480018889520375895058839125174559994890 \\
18 & 0.04480018889520375912411377530928727493355 &
0.04480018889520375909792228228963030071768 \\
19 & 0.04480018889520375910184581646734771612385 &
0.04480018889520375910126228938039327226238 \\
20 & 0.04480018889520375910134847750070616671897 &
0.04480018889520375910133583118877440345999 \\
21 & 0.04480018889520375910133767505507731700176 &
0.04480018889520375910133740784342599125332 \\
22 & 0.04480018889520375910133744634213765970226 &
0.04480018889520375910133744082644719944017 \\
23 & 0.04480018889520375910133744161242203858736 &
0.04480018889520375910133744150100303657243 \\
24 & 0.04480018889520375910133744151671873083494 &
0.04480018889520375910133744151451271412294 \\
25 & 0.04480018889520375910133744151482093207536 &
0.04480018889520375910133744151477806241713 \\
26 & 0.04480018889520375910133744151478399919048 &
0.04480018889520375910133744151478318049710 \\
27 & 0.04480018889520375910133744151478329293722 &
0.04480018889520375910133744151478327755544 \\
28 & 0.04480018889520375910133744151478327965163 &
0.04480018889520375910133744151478327936702 \\
29 & 0.04480018889520375910133744151478327940553 &
0.04480018889520375910133744151478327940034 \\
30 & 0.04480018889520375910133744151478327940103 &
0.04480018889520375910133744151478327940094 \\ \hline
\end{tabular}
}
\end{center}
\par
{\tiny \ }
\end{table}

\begin{table}[H]
\caption{Upper and lower bounds to the critical parameter $g_6$}
\label{tab:g6}{\tiny {\ }}
\par
\begin{center}
{\tiny
\begin{tabular}{rll}
\hline
\multicolumn{1}{c}{$D$} & \multicolumn{1}{c}{$d=0$} & \multicolumn{1}{c}{$d=1
$} \\ \hline
9 & 0.03555789093100854947852001233906087375285 &
0.03092572893772660886926073199814728234093 \\
10 & 0.03161133560525730782271008444032283771361 &
0.03145849368669941826818885316031905637790 \\
11 & 0.03149011870542296376338502841225772774296 &
0.03148364917358371728205366708630422335728 \\
12 & 0.03148494329367438817018069366727460457784 &
0.03148468923680296515244178265796186430402 \\
13 & 0.03148473822437121499799204350554556202120 &
0.03148472893577904807755520184965076889074 \\
14 & 0.03148473066939209541383045474594042161208 &
0.03148473035061312995068282335883952417254 \\
15 & 0.03148473040841264012200055796453306251009 &
0.03148473039807105133151396352293524948923 \\
16 & 0.03148473039989824041338340929593557562405 &
0.03148473039957923859546643481029290114944 \\
17 & 0.03148473039963430349984031102576795682861 &
0.03148473039962490047024519342698839474870 \\
18 & 0.03148473039962648972467472714013522248438 &
0.03148473039962622373951281024412934796390 \\
19 & 0.03148473039962626784077105845026744982718 &
0.03148473039962626059387043871581476604155 \\
20 & 0.03148473039962626177453805053554733260776 &
0.03148473039962626158375823527601914981292 \\
21 & 0.03148473039962626161434349129423137002648 &
0.03148473039962626160947714346266678266061 \\
22 & 0.03148473039962626161024580222497003617042 &
0.03148473039962626161012523650616399469066 \\
23 & 0.03148473039962626161014402047172021513590 &
0.03148473039962626161014111289775964461238 \\
24 & 0.03148473039962626161014156015270460639019 &
0.03148473039962626161014149176851390225706 \\
25 & 0.03148473039962626161014150216346828657935 &
0.03148473039962626161014150059223332966340 \\
26 & 0.03148473039962626161014150082844005299243 &
0.03148473039962626161014150079311744772934 \\
27 & 0.03148473039962626161014150079837272301895 &
0.03148473039962626161014150079759470472809 \\
28 & 0.03148473039962626161014150079770933562803 &
0.03148473039962626161014150079769252461137 \\
29 & 0.03148473039962626161014150079769497890499 &
0.03148473039962626161014150079769462216118 \\
30 & 0.03148473039962626161014150079769467379548 &
0.03148473039962626161014150079769466635289 \\ \hline
\end{tabular}
}
\end{center}
\par
{\tiny \ }
\end{table}

\begin{table}[H]
\caption{Upper and lower bounds to the critical parameter $g_8$}
\label{tab:g8}{\tiny {\ }}
\par
\begin{center}
{\tiny
\begin{tabular}{rll}
\hline
\multicolumn{1}{c}{$D$} & \multicolumn{1}{c}{$d=0$} & \multicolumn{1}{c}{$d=1
$} \\ \hline
12 & 0.02454606377413199119828689007811326833190 &
0.02421917801850957375955340047328637544178 \\
13 & 0.02428661490509848495426723668328039541065 &
0.02427219192148520285307831828202703716164 \\
14 & 0.02427518920350644058599111753338208200674 &
0.02427457554899199037037423016257647520912 \\
15 & 0.02427469912043926734852197785524518494693 &
0.02427467461304566168142545908666649680638 \\
16 & 0.02427467940328849131593270483231916087025 &
0.02427467847982376530194467187325031595840 \\
17 & 0.02427467865552462797424190681519693334287 &
0.02427467862251144116586540572557279217836 \\
18 & 0.02427467862864074047936428296833975349799 &
0.02427467862751567792228058363500195486763 \\
19 & 0.02427467862771994664763439199207617680632 &
0.02427467862768324470768285411390709785292 \\
20 & 0.02427467862768977342041040780145606689484 &
0.02427467862768862316320148384898791864478 \\
21 & 0.02427467862768882396042257443841404567986 &
0.02427467862768878921703836739073112772072 \\
22 & 0.02427467862768879517757388028203141313335 &
0.02427467862768879416334573266611758647620 \\
23 & 0.02427467862768879433456546695526061746882 &
0.02427467862768879430587992894599527275900 \\
24 & 0.02427467862768879431065060211425612673539 &
0.02427467862768879430986280571665336624900 \\
25 & 0.02427467862768879430999200803802694651858 &
0.02427467862768879430997095825833740269428 \\
26 & 0.02427467862768879430997436578735672832845 &
0.02427467862768879430997381759182836736611 \\
27 & 0.02427467862768879430997390525558585568438 &
0.02427467862768879430997389131843556420278 \\
28 & 0.02427467862768879430997389352173737221345 &
0.02427467862768879430997389317532507421902 \\
29 & 0.02427467862768879430997389322950042565877 &
0.02427467862768879430997389322107162171259 \\
30 & 0.02427467862768879430997389322237643876678 &
0.02427467862768879430997389322217543063241 \\ \hline
\end{tabular}
}
\end{center}
\par
{\tiny \ }
\end{table}

\begin{table}[H]
\caption{Upper and lower bounds to the critical parameter $g_{10}$}
\label{tab:g10}{\tiny {\ }}
\par
\begin{center}
{\tiny
\begin{tabular}{rll}
\hline
\multicolumn{1}{c}{$D$} & \multicolumn{1}{c}{$d=0$} & \multicolumn{1}{c}{$d=1
$} \\ \hline
14 & 0.02047221602266282751456657741834448703861 &
0.01963359120980542253748626901810922604369 \\
15 & 0.01977995619382549576732985165234401774013 &
0.01974693950909705411949585984494718255830 \\
16 & 0.01975395366584947117993371023280556305747 &
0.01975247206868859784344760242869068628597 \\
17 & 0.01975277987117262166852573265229226959880 &
0.01975271678710470678106421086781565332383 \\
18 & 0.01975272954367646835097927623308544993621 &
0.01975272699679626485875877128889725490927 \\
19 & 0.01975272749910508026589732684965854046262 &
0.01975272740119146954343985854466858794848 \\
20 & 0.01975272742006407830784338141483236540174 &
0.01975272741646549344162513054957393198253 \\
21 & 0.01975272741714457999855452711001915380620 &
0.01975272741701770267830233753828310016295 \\
22 & 0.01975272741704118117756836119192878870243 &
0.01975272741703687653629231814017786830575 \\
23 & 0.01975272741703765875505103202559254656262 &
0.01975272741703751783247422983584944371101 \\
24 & 0.01975272741703754301058620864013871701965 &
0.01975272741703753854808514709255666516573 \\
25 & 0.01975272741703753933289078549121445588845 &
0.01975272741703753919590212025527989701806 \\
26 & 0.01975272741703753921964046956525127252133 &
0.01975272741703753921555577820644864641499 \\
27 & 0.01975272741703753921625385870378559062189 &
0.01975272741703753921613534202517155387932 \\
28 & 0.01975272741703753921615533452549872575039 &
0.01975272741703753921615198293899204352528 \\
29 & 0.01975272741703753921615254141912901863930 &
0.01975272741703753921615244890377596374998 \\
30 & 0.01975272741703753921615246414220476261028 &
0.01975272741703753921615246164614317022386 \\ \hline
\end{tabular}
}
\end{center}
\par
{\tiny \ }
\end{table}

\begin{table}[H]
\caption{Upper and lower bounds to the critical parameter $g_1$}
\label{tab:g1}{\tiny {\ }}
\par
\begin{center}
{\tiny
\begin{tabular}{rll}
\hline
\multicolumn{1}{c}{$D$} & \multicolumn{1}{c}{$d=0$} & \multicolumn{1}{c}{$d=1
$} \\ \hline
2 & 0.2086996778999803716648562090815857819745 &
0.1631480430486902406431149910168791819850 \\
3 & 0.1718969800128251621850466269627633199548 &
0.1703607047409994702758081751158201171033 \\
4 & 0.1706205139560892552710851925000027130523 &
0.1705785830384682823330294547098227728383 \\
5 & 0.1705851234732303691401645911164046595938 &
0.1705841331044978946845395959683158081996 \\
6 & 0.1705842793540777621350223723941786562760 &
0.1705842582207720111749295014232036857427 \\
7 & 0.1705842612172584531052118710606141112861 &
0.1705842607994262597452448410639135511219 \\
8 & 0.1705842608568300570962041004073058923890 &
0.1705842608490478764194508397061227525251 \\
9 & 0.1705842608500903205860724454343000132200 &
0.1705842608499521929400533330273499936642 \\
10 & 0.1705842608499703147949847504660105310099 &
0.1705842608499679587616751011174176672218 \\
11 & 0.1705842608499682625237635665875689738621 &
0.1705842608499682236606681996943088012139 \\
12 & 0.1705842608499682285973929807899222041968 &
0.1705842608499682279744377896817537199688 \\
13 & 0.1705842608499682280525621246154707680480 &
0.1705842608499682280428211446751570550038 \\
14 & 0.1705842608499682280440291352128456950145 &
0.1705842608499682280438800921911945957748 \\
15 & 0.1705842608499682280438983932598809251813 &
0.1705842608499682280438961562092488713937 \\
16 & 0.1705842608499682280438964284902789575884 &
0.1705842608499682280438963954837566117108 \\
17 & 0.1705842608499682280438963994695384958829 &
0.1705842608499682280438963989899794778859 \\
18 & 0.1705842608499682280438963990474787501195 &
0.1705842608499682280438963990406073441846 \\
19 & 0.1705842608499682280438963990414259180386 &
0.1705842608499682280438963990413286969484 \\
20 & 0.1705842608499682280438963990413402105333 &
0.1705842608499682280438963990413388507761 \\
21 & 0.1705842608499682280438963990413390109393 &
0.1705842608499682280438963990413389921218 \\
22 & 0.1705842608499682280438963990413389943273 &
0.1705842608499682280438963990413389940694 \\
23 & 0.1705842608499682280438963990413389940995 &
0.1705842608499682280438963990413389940960 \\
24 & 0.1705842608499682280438963990413389940964 &
0.1705842608499682280438963990413389940963 \\
25 & 0.1705842608499682280438963990413389940963 &
0.1705842608499682280438963990413389940963 \\
26 & 0.1705842608499682280438963990413389940963 &
0.1705842608499682280438963990413389940963 \\ \hline
\end{tabular}
}
\end{center}
\par
{\tiny \ }
\end{table}

\begin{table}[H]
\caption{Upper and lower bounds to the critical parameter $g_3$}
\label{tab:g3}{\tiny {\ }}
\par
\begin{center}
{\tiny
\begin{tabular}{rll}
\hline
\multicolumn{1}{c}{$D$} & \multicolumn{1}{c}{$d=0$} & \multicolumn{1}{c}{$d=1
$} \\ \hline
5 & 0.06658423997291568499134684206889912226812 &
0.06517643978579263140844752299484901358385 \\
6 & 0.06545525969547331631663568171916144298109 &
0.06540187408173481594068149565108956282900 \\
7 & 0.06541178093769876825737277164051845877719 &
0.06540999426358348159542255715630073839591 \\
8 & 0.06541030843742374541892381357813324610058 &
0.06541025444317428244130444366902325905977 \\
9 & 0.06541026353211840459137158434656862367146 &
0.06541026203083056984744763345599166019452 \\
10 & 0.06541026227455184719285856903449364470416 &
0.06541026223561071297259902248086393032143 \\
11 & 0.06541026224174188141376880992617857852887 &
0.06541026224078959072891751291251930008276 \\
12 & 0.06541026224093564261132742105590729066078 &
0.06541026224091350499126239375145555001202 \\
13 & 0.06541026224091682376029353614954016102345 &
0.06541026224091633132610343955357326924266 \\
14 & 0.06541026224091640368935822734476921154305 &
0.06541026224091639315192254921567203670695 \\
15 & 0.06541026224091639467325676065672782241449 &
0.06541026224091639445538878773831124637217 \\
16 & 0.06541026224091639448635085915507961920407 &
0.06541026224091639448198260617462957942654 \\
17 & 0.06541026224091639448259465520064910032909 &
0.06541026224091639448250946052769307853630 \\
18 & 0.06541026224091639448252124533611449529231 &
0.06541026224091639448251962486031173960277 \\
19 & 0.06541026224091639448251984641940638968848 &
0.06541026224091639448251981629125127661268 \\
20 & 0.06541026224091639448251982036686428666309 &
0.06541026224091639448251981981827720688257 \\
21 & 0.06541026224091639448251981989176565853991 &
0.06541026224091639448251981988196632469101 \\
22 & 0.06541026224091639448251981988326726046886 &
0.06541026224091639448251981988309528353080 \\
23 & 0.06541026224091639448251981988311792520262 &
0.06541026224091639448251981988311495602909 \\
24 & 0.06541026224091639448251981988311534392275 &
0.06541026224091639448251981988311529343365 \\
25 & 0.06541026224091639448251981988311529998220 &
0.06541026224091639448251981988311529913574 \\
26 & 0.06541026224091639448251981988311529924479 &
0.06541026224091639448251981988311529923079 \\
27 & 0.06541026224091639448251981988311529923258 &
0.06541026224091639448251981988311529923235 \\
28 & 0.06541026224091639448251981988311529923238 &
0.06541026224091639448251981988311529923238 \\
29 & 0.06541026224091639448251981988311529923238 &
0.06541026224091639448251981988311529923238 \\ \hline
\end{tabular}
}
\end{center}
\par
{\tiny \ }
\end{table}

\begin{table}[H]
\caption{Upper and lower bounds to the critical parameter $g_5$}
\label{tab:g5}{\tiny {\ }}
\par
\begin{center}
{\tiny
\begin{tabular}{rll}
\hline
\multicolumn{1}{c}{$D$} & \multicolumn{1}{c}{$d=0$} & \multicolumn{1}{c}{$d=1
$} \\ \hline
7 & 0.04235342681449874828379645588296849560493 &
0.04008468238431420706503999644754864395157 \\
8 & 0.04053492929116994535350218365600472273380 &
0.04044086981225583569490363708593431083698 \\
9 & 0.04045975479435392023038061859624314153650 &
0.04045604655295448418017316688034199143398 \\
10 & 0.04045675821089881525694159182373550765977 &
0.04045662444278018054848586537819408671823 \\
11 & 0.04045664910617543504412152668595151889447 &
0.04045664463966676394297232250175238597000 \\
12 & 0.04045664543514700963483422664255168886621 &
0.04045664529566647699538229268938359251518 \\
13 & 0.04045664531976836367316867234905644507400 &
0.04045664531566035784685046592826441898386 \\
14 & 0.04045664531635156317423426766344794710382 &
0.04045664531623666821495891297369542396561 \\
15 & 0.04045664531625554841934268920858151245483 &
0.04045664531625247944390720501249170050884 \\
16 & 0.04045664531625297319525699655560538570218 &
0.04045664531625289453074904143020885598121 \\
17 & 0.04045664531625290694769780460302975452153 &
0.04045664531625290500497641557666478625810 \\
18 & 0.04045664531625290530637799647025575437974 &
0.04045664531625290525999200685692030298506 \\
19 & 0.04045664531625290526707618855393056265051 &
0.04045664531625290526600220208649305693245 \\
20 & 0.04045664531625290526616387945812489665168 &
0.04045664531625290526613970441442199842509 \\
21 & 0.04045664531625290526614329587540741524176 &
0.04045664531625290526614276563623192474136 \\
22 & 0.04045664531625290526614284345304123226822 &
0.04045664531625290526614283209838581067494 \\
23 & 0.04045664531625290526614283374603863631471 &
0.04045664531625290526614283350822558832505 \\
24 & 0.04045664531625290526614283354237364763735 &
0.04045664531625290526614283353749460735177 \\
25 & 0.04045664531625290526614283353818837539638 &
0.04045664531625290526614283353809018357822 \\
26 & 0.04045664531625290526614283353810401871459 &
0.04045664531625290526614283353810207782254 \\
27 & 0.04045664531625290526614283353810234895946 &
0.04045664531625290526614283353810231123684 \\
28 & 0.04045664531625290526614283353810231646435 &
0.04045664531625290526614283353810231574271 \\
29 & 0.04045664531625290526614283353810231584196 &
0.04045664531625290526614283353810231582836 \\
30 & 0.04045664531625290526614283353810231583022 &
0.04045664531625290526614283353810231582996 \\ \hline
\end{tabular}
}
\end{center}
\par
{\tiny \ }
\end{table}

\begin{table}[H]
\caption{Upper and lower bounds to the critical parameter $g_7$}
\label{tab:g7}{\tiny {\ }}
\par
\begin{center}
{\tiny
\begin{tabular}{rll}
\hline
\multicolumn{1}{c}{$D$} & \multicolumn{1}{c}{$d=0$} & \multicolumn{1}{c}{$d=1
$} \\ \hline
10 & 0.02944043545506540844767350198035741685093 &
0.02925382514375896464671235971372483554623 \\
11 & 0.02929263011019324957778320977958604205969 &
0.02928459515933800353350736998850181395998 \\
12 & 0.02928622095818320661586489613547243254370 &
0.02928589785226081725795933860791764382483 \\
13 & 0.02928596094712212155460236641440229752371 &
0.02928594882668895559114675244822546899524 \\
14 & 0.02928595111921799816826053773236604219610 &
0.02928595069189246851138552618063367657143 \\
15 & 0.02928595077045074072266816097994054501954 &
0.02928595075619700683165712945846651860227 \\
16 & 0.02928595075875122972135286911608051280956 &
0.02928595075829890538280596210753436363205 \\
17 & 0.02928595075837810986295798740066779417270 &
0.02928595075836438880075684513209985568635 \\
18 & 0.02928595075836674155819019733965787007034 &
0.02928595075836634205850095626467616762134 \\
19 & 0.02928595075836640926179834040441440539466 &
0.02928595075836639805781956766424063211803 \\
20 & 0.02928595075836639990974309652901026078448 &
0.02928595075836639960614840049473166735312 \\
21 & 0.02928595075836639965552572641176616526079 &
0.02928595075836639964755575998040863415492 \\
22 & 0.02928595075836639964883280531828487269726 &
0.02928595075836639964862961870287410535980 \\
23 & 0.02928595075836639964866172824801636371768 &
0.02928595075836639964865668711007929634103 \\
24 & 0.02928595075836639964865747357201094142001 &
0.02928595075836639964865735162377861613877 \\
25 & 0.02928595075836639964865737042175116707099 &
0.02928595075836639964865736754058295358149 \\
26 & 0.02928595075836639964865736797974866929830 &
0.02928595075836639964865736791316517905794 \\
27 & 0.02928595075836639964865736792320799178975 &
0.02928595075836639964865736792170081663888 \\
28 & 0.02928595075836639964865736792192590717946 &
0.02928595075836639964865736792189244929585 \\
29 & 0.02928595075836639964865736792189739977306 &
0.02928595075836639964865736792189667055062 \\
30 & 0.02928595075836639964865736792189677750333 &
0.02928595075836639964865736792189676188285 \\ \hline
\end{tabular}
}
\end{center}
\par
{\tiny \ }
\end{table}

\begin{table}[H]
\caption{Upper and lower bounds to the critical parameter $g_9$}
\label{tab:g9}{\tiny {\ }}
\par
\begin{center}
{\tiny
\begin{tabular}{rll}
\hline
\multicolumn{1}{c}{$D$} & \multicolumn{1}{c}{$d=0$} & \multicolumn{1}{c}{$d=1
$} \\ \hline
12 & 0.02329060717345882812929809993469954354839 &
0.02288168656122881606834380073412630921766 \\
13 & 0.02296466585093838369363438315345906760748 &
0.02294670804722385869383190696061532989902 \\
14 & 0.02295046689463449122205871968561471535248 &
0.02294969054566492819024936034224687544502 \\
15 & 0.02294984828211896691375200728655512450953 &
0.02294981670571586685777002151656828297988 \\
16 & 0.02294982293735504571626453992935772109552 &
0.02294982172407890215326351386560285988779 \\
17 & 0.02294982195726747110669443863071991252322 &
0.02294982191299831900377208175041359575596 \\
18 & 0.02294982192130408334069325371494219260989 &
0.02294982191976321787820458223173065952988 \\
19 & 0.02294982192004600800576202711717634586091 &
0.02294982191999464238368090308647173457388 \\
20 & 0.02294982192000388025314759124176339867059 &
0.02294982192000223462001097458990789510519 \\
21 & 0.02294982192000252510040859515776395449364 &
0.02294982192000247427575548992663346212595 \\
22 & 0.02294982192000248309331156982426267954702 &
0.02294982192000248157600760241986117382833 \\
23 & 0.02294982192000248183505209869665923986167 &
0.02294982192000248179116130013135788646917 \\
24 & 0.02294982192000248179854345959064375280489 &
0.02294982192000248179731060544765190386249 \\
25 & 0.02294982192000248179751508988595278496814 &
0.02294982192000248179748139795654468872600 \\
26 & 0.02294982192000248179748691365471395107790 &
0.02294982192000248179748601628477222654560 \\
27 & 0.02294982192000248179748616140306635176449 &
0.02294982192000248179748613807215578726825 \\
28 & 0.02294982192000248179748614180187080893006 &
0.02294982192000248179748614120890980039698 \\
29 & 0.02294982192000248179748614130267722344290 &
0.02294982192000248179748614128792630866716 \\
30 & 0.02294982192000248179748614129023512798250 &
0.02294982192000248179748614128987552346398 \\ \hline
\end{tabular}
}
\end{center}
\par
{\tiny \ }
\end{table}

\begin{table}[H]
\caption{Upper and lower bounds to the critical parameter $g_{11}$}
\label{tab:g11}{\tiny {\ }}
\par
\begin{center}
{\tiny
\begin{tabular}{rll}
\hline
\multicolumn{1}{c}{$D$} & \multicolumn{1}{c}{$d=0$} & \multicolumn{1}{c}{$d=1
$} \\ \hline
14 & 0.01996781952733978066990651885169494601712 &
0.01872252524814441498777702067663915812970 \\
15 & 0.01890198638848331888258042805590161397986 &
0.01886069386249469343558391653645193592948 \\
16 & 0.01886948779463646969880275156276806629728 &
0.01886761704220707966640143390518547691489 \\
17 & 0.01886800824507166015980470286736442365237 &
0.01886792750167288410437554271348871817449 \\
18 & 0.01886794394845459026863114135387378164300 &
0.01886794064000340464095534517769369533121 \\
19 & 0.01886794129758633264365840899606626347236 &
0.01886794116838374607914916662761523751379 \\
20 & 0.01886794119349012426038980610674152516619 &
0.01886794118866314741294671844248716274723 \\
21 & 0.01886794118958173342426775990842042714033 &
0.01886794118940863900110547005121626575133 \\
22 & 0.01886794118944094780343142406309400748702 &
0.01886794118943497221768172831049899983764 \\
23 & 0.01886794118943606768966856387448227211230 &
0.01886794118943586856954280873755757272623 \\
24 & 0.01886794118943590446563326063745363934179 &
0.01886794118943589804591666167271857500859 \\
25 & 0.01886794118943589918520561482923202470277 &
0.01886794118943589898452368838333550424238 \\
26 & 0.01886794118943589901961832711799440636334 &
0.01886794118943589901352396387709244724591 \\
27 & 0.01886794118943589901457511041947361581581 &
0.01886794118943589901439500217258055563250 \\
28 & 0.01886794118943589901442566578641554737888 &
0.01886794118943589901442047763434765401488 \\
29 & 0.01886794118943589901442135016026667735164 &
0.01886794118943589901442120428100247967571 \\
30 & 0.01886794118943589901442122853193989182105 &
0.01886794118943589901442122452280735553630 \\ \hline
\end{tabular}
}
\end{center}
\par
{\tiny \ }
\end{table}

\begin{table}[H]
\caption{Eigenvalue $E_0$ of $V(x)=-20x^2+2x^4$}
\label{tab:e0}{\tiny {\ }}
\par
\begin{center}
{\tiny
\begin{tabular}{rll}
\hline
\multicolumn{1}{c}{$D$} & \multicolumn{1}{c}{$d=0$} & \multicolumn{1}{c}{$d=1
$} \\ \hline
6 & -45.48176505716663623746030853874339294793 &
-43.31505048121806269799121924974007323963 \\
7 & -43.93562409115915060703569782152785395376 &
-43.73294969603763128167544575228979847158 \\
8 & -43.79290846027861844357310305720444485022 &
-43.77551108808928229026247019926137805250 \\
9 & -43.78034879437027161346303655319189662654 &
-43.77904558645083988975394614271026643160 \\
10 & -43.77938563684300145774048474622395679774 &
-43.77929944424936518397005859939559664346 \\
11 & -43.77932070326800389143960331807526523524 &
-43.77931559176863118190405961154233053389 \\
12 & -43.77931679174058833577150015312372370423 &
-43.77931651628984503218700608762334214196 \\
13 & -43.77931657819722294399860448861432496436 &
-43.77931656455787726577858851166259130381 \\
14 & -43.77931656750691564181875020351006669623 &
-43.77931656688052264492602808929547704601 \\
15 & -43.77931656701135087269142632700246175155 &
-43.77931656698445901112942619108928218250 \\
16 & -43.77931656698990345125782055132247261168 &
-43.77931656698881697367559286304162915467 \\
17 & -43.77931656698903083010421439365129825553 &
-43.77931656698898928351091605403643087867 \\
18 & -43.77931656698899725461293104654246458294 &
-43.77931656698899574344803837551937305507 \\
19 & -43.77931656698899602667888879332542581751 &
-43.77931656698899597417221606078658466804 \\
20 & -43.77931656698899598380452284334618538122 &
-43.77931656698899598205519055854216039383 \\
21 & -43.77931656698899598236983028579062162137 &
-43.77931656698899598231376207477055682696 \\
22 & -43.77931656698899598232366434613920385196 &
-43.77931656698899598232193049349880226816 \\
23 & -43.77931656698899598232223157555255010996 &
-43.77931656698899598232217970971274714616 \\
24 & -43.77931656698899598232218857562333915175 &
-43.77931656698899598232218707135367394310 \\
25 & -43.77931656698899598232218732474833030185 &
-43.77931656698899598232218728236042310832 \\
26 & -43.77931656698899598232218728940341828025 &
-43.77931656698899598232218728824080056198 \\
27 & -43.77931656698899598232218728843150872694 &
-43.77931656698899598232218728840041742956 \\
28 & -43.77931656698899598232218728840545621676 &
-43.77931656698899598232218728840464430755 \\
29 & -43.77931656698899598232218728840477440128 &
-43.77931656698899598232218728840475366921 \\
30 & -43.77931656698899598232218728840475695571 &
-43.77931656698899598232218728840475643739 \\ \hline
\end{tabular}
}
\end{center}
\par
{\tiny \ }
\end{table}

\begin{table}[H]
\caption{Eigenvalue $E_1$ of $V(x)=-20x^2+2x^4$}
\label{tab:e1}{\tiny {\ }}
\par
\begin{center}
{\tiny
\begin{tabular}{rll}
\hline
\multicolumn{1}{c}{$D$} & \multicolumn{1}{c}{$d=0$} & \multicolumn{1}{c}{$d=1
$} \\ \hline
6 & -44.28978145133939862149419754620613346698 &
-43.62795163389018669860165633098424087735 \\
7 & -43.82616673070973173577840556883805307921 &
-43.76576795931880241521357215234857485653 \\
8 & -43.78312553803835348336577354919495769796 &
-43.77828451010011593274177269922005910433 \\
9 & -43.77958746079137554333463082137428876952 &
-43.77924739174091404787515603368964260950 \\
10 & -43.77933358307886826118081591082050673091 &
-43.77931232397705691808359150077738077252 \\
11 & -43.77931743547205578754257366879262383436 &
-43.77931623549969136036517949635666556850 \\
12 & -43.77931651095044783505835230224087611246 &
-43.77931644904306266813723614415863395298 \\
13 & -43.77931646268240972489009270135986067432 &
-43.77931645973337103998145392160130025351 \\
14 & -43.77931646035976410196849335890222640688 &
-43.77931646022893586058420780053698848631 \\
15 & -43.77931646025582772494455556092372338615 &
-43.77931646025038328424957290289598406600 \\
16 & -43.77931646025146976194486559104801805466 &
-43.77931646025125590549398887865761953113 \\
17 & -43.77931646025129745209161080295437763140 &
-43.77931646025128948098876629023213347179 \\
18 & -43.77931646025129099215381622204479194084 &
-43.77931646025129070892293632955493647256 \\
19 & -43.77931646025129076142961452624986298875 &
-43.77931646025129075179730674129523337257 \\
20 & -43.77931646025129075354663920814516135221 &
-43.77931646025129075323199944815342215392 \\
21 & -43.77931646025129075328806766500827785286 &
-43.77931646025129075327816539260914168142 \\
22 & -43.77931646025129075327989924542997826452 &
-43.77931646025129075327959816334489803674 \\
23 & -43.77931646025129075327965002919009846775 &
-43.77931646025129075327964116327858382287 \\
24 & -43.77931646025129075327964266754840557475 &
-43.77931646025129075327964241415372284625 \\
25 & -43.77931646025129075327964245654163445092 &
-43.77931646025129075327964244949863854605 \\
26 & -43.77931646025129075327964245066125638531 &
-43.77931646025129075327964245047054820050 \\
27 & -43.77931646025129075327964245050163950112 &
-43.77931646025129075327964245049660071339 \\
28 & -43.77931646025129075327964245049741262269 &
-43.77931646025129075327964245049728252895 \\
29 & -43.77931646025129075327964245049730326102 &
-43.77931646025129075327964245049729997452 \\
30 & -43.77931646025129075327964245049730049283 &
-43.77931646025129075327964245049730041150 \\ \hline
\end{tabular}
}
\end{center}
\par
{\tiny \ }
\end{table}

\begin{table}[H]
\caption{Eigenvalue of the double well $V(x)=-20x^2+2x^4$ from expansion
about one of the minima}
\label{tab:e_ns}{\tiny {\ }}
\par
\begin{center}
{\tiny
\begin{tabular}{rll}
\hline
\multicolumn{1}{c}{$D$} & \multicolumn{1}{c}{$E$} & \multicolumn{1}{c}{$f_0$}
\\ \hline
\multicolumn{3}{c}{$d=0$} \\
3 & -43.77931669031571738751481356566038039124 &
0.2273917394440904379752945788994227972398 \\
4 & -43.77931651361968401280584765608342823829 &
0.2273916971399277265908076160133509098932 \\
5 & -43.77931651362014038166816249872705138561 &
0.2273916971400760682068043142501069051802 \\
6 & -43.77931651362014024495550010192866031029 &
0.2273916971400760353994260712747814338263 \\
7 & -43.77931651362014498836136885879631004559 &
0.2273916971400775105333097837863779501094 \\
8 & -43.77931651362014510834900235709469809467 &
0.2273916971400775478952431605722187981473 \\
9 & -43.77931651362014514928353456063329997293 &
0.2273916971400775606506174073218214534183 \\
10 & -43.77931651362014009917520045377859851074 &
0.2273916971400759871019196828844221848766 \\
11 & -43.77931651362014498790325227778299494589 &
0.2273916971400775103665439032195059408009 \\
\multicolumn{3}{c}{$d=1$} \\
3 & -43.77931651361612431420607712081719445236 &
0.2273916971387571156369644228290555283616 \\
4 & -43.77931651362012751710629596872829894209 &
0.2273916971400720236355367680185887323197 \\
5 & -43.77931651362014416185777976622515513944 &
0.2273916971400772528329334561526339961353 \\
6 & -43.77931651362014167227391733804968489512 &
0.2273916971400764788942388256767738879596* \\
7 & -43.77931651362014410032754942613776847221 &
0.2273916971400772338084116146065912034121 \\
8 & -43.77931651362014339248155845283235823475 &
0.2273916971400770132576683468221320558258 \\
9 & -43.77931651362014447117488104583915749563 &
0.2273916971400773493605499071994360392614 \\
10 & -43.77931651362014400180312517086275181785 &
0.2273916971400772031103818128168790570476 \\ \hline
\end{tabular}
}
\end{center}
\par
{\tiny \ }
\end{table}

\begin{table}[H]
\caption{Lowest resonance for $V(x)=x^2-0.1 x^4$}
\label{tab:e_res}{\tiny {\ }}
\par
\begin{center}
{\tiny
\begin{tabular}{rll}
\hline \multicolumn{1}{c}{$D$} & \multicolumn{1}{c}{$\Re E$} &
\multicolumn{1}{c}{$\Im E$} \\ \hline
\multicolumn{3}{c}{$d=0$} \\
3 & 0.9009045987434388961113826558408929050585 &
0.006531275248132131969937339072308577550367 \\
4 & 0.9006785761554979170185849838170952561575 &
0.006697172917429490409114415779897272309962 \\
5 & 0.9006728969139677462512941148232908174675 &
0.006693419931424615406839099221196299082930 \\
6 & 0.9006729020396943939429810890309258597079 &
0.006693282383360464347154292684069904696091 \\
7 & 0.9006729040491209448507241419711280157202 &
0.006693280869143828507690412010788048372771 \\
8 & 0.9006729040915603251405575905226281082941 &
0.006693280875265106991770610308683065847521 \\
9 & 0.9006729040920144630597265124954551335021 &
0.006693280875789241683725707577273885929508 \\
10 & 0.9006729040920151108192341764268998578420 &
0.006693280875799990838393406602497907148018 \\
11 & 0.9006729040920150269574289739136914578854 &
0.006693280875800129218078648202714592941434 \\
12 & 0.9006729040920150248388637300439191244513 &
0.006693280875800130271842860245002534919132 \\
13 & 0.9006729040920150248051243749312897357163 &
0.006693280875800130269532817035582731535822 \\
14 & 0.9006729040920150248047249847012756191763 &
0.006693280875800130269277606514629002483066 \\
15 & 0.9006729040920150248047216955367124563283 &
0.006693280875800130269271964600663717648139 \\
16 & 0.9006729040920150248047216887862151349051 &
0.006693280875800130269271876206213583967297 \\
17 & 0.9006729040920150248047216891991689220470 &
0.006693280875800130269271875092705208083531 \\
18 & 0.9006729040920150248047216892100957991600 &
0.006693280875800130269271875081399548665397 \\
19 & 0.9006729040920150248047216892102850542174 &
0.006693280875800130269271875081318291974165 \\
20 & 0.9006729040920150248047216892102877256260 &
0.006693280875800130269271875081318229210992 \\
21 & 0.9006729040920150248047216892102877579668 &
0.006693280875800130269271875081318240835207 \\
22 & 0.9006729040920150248047216892102877583018 &
0.006693280875800130269271875081318241118075 \\
23 & 0.9006729040920150248047216892102877583046 &
0.006693280875800130269271875081318241122880 \\
24 & 0.9006729040920150248047216892102877583046 &
0.006693280875800130269271875081318241122949 \\
25 & 0.9006729040920150248047216892102877583046 &
0.006693280875800130269271875081318241122950 \\
26 & 0.9006729040920150248047216892102877583046 &
0.006693280875800130269271875081318241122950 \\
\multicolumn{3}{c}{$d=1$} \\
3 & 0.9006269386934407149024188879619797335942 &
0.006692165141048608819080712012111923978095 \\
4 & 0.9006724819314438873902357659829302302388 &
0.006692380922317620997669150541899710733084 \\
5 & 0.9006729133557364622189247126467344851372 &
0.006693264258310679158403839879649976013988 \\
6 & 0.9006729044182223618433366709225523876453 &
0.006693280799696702428168425080430007058535 \\
7 & 0.9006729040968377563374267222635288479840 &
0.006693280878551503991618828166689870608647 \\
8 & 0.9006729040920469156664148684004597890937 &
0.006693280875882034788729653473581427028750 \\
9 & 0.9006729040920146849762905383362938705186 &
0.006693280875801427980509583740830419378055 \\
10 & 0.9006729040920150100162224583175926763758 &
0.006693280875800143543938942464596749680557 \\
11 & 0.9006729040920150245224972608124165923811 &
0.006693280875800130322255056366508536933007 \\
12 & 0.9006729040920150248008764220038503344438 &
0.006693280875800130267961175438433627644701 \\
13 & 0.9006729040920150248046829813016628536234 &
0.006693280875800130269230805638311673398573 \\
14 & 0.9006729040920150248047214659649977703386 &
0.006693280875800130269271136460543267193167 \\
15 & 0.9006729040920150248047216907377831922129 &
0.006693280875800130269271864791037012412317 \\
16 & 0.9006729040920150248047216892849347231816 &
0.006693280875800130269271874964638556109025 \\
17 & 0.9006729040920150248047216892117960284237 &
0.006693280875800130269271875080292909149647 \\
18 & 0.9006729040920150248047216892103110181208 &
0.006693280875800130269271875081313332926222 \\
19 & 0.9006729040920150248047216892102880610630 &
0.006693280875800130269271875081318292154781 \\
20 & 0.9006729040920150248047216892102877616995 &
0.006693280875800130269271875081318243107606 \\
21 & 0.9006729040920150248047216892102877583365 &
0.006693280875800130269271875081318241161447 \\
22 & 0.9006729040920150248047216892102877583048 &
0.006693280875800130269271875081318241123542 \\
23 & 0.9006729040920150248047216892102877583046 &
0.006693280875800130269271875081318241122958 \\
24 & 0.9006729040920150248047216892102877583046 &
0.006693280875800130269271875081318241122950 \\
25 & 0.9006729040920150248047216892102877583046 &
0.006693280875800130269271875081318241122950 \\ \hline
\end{tabular}
}
\end{center}
\par
{\tiny \ }
\end{table}


\begin{thebibliography}{9}
\bibitem{T05}  A. V. Turbiner, Anharmonic oscillator and double-well
potential: approximating eigenfunctions, Lett. Math. Phys. 74 (2005) 169-180.

\bibitem{T10}  A. V. Turbiner, Double well potential: perturbation theory,
tunneling, WKB (beyond instantons), Int. J. Mod. Phys. A 25 (2010) 647-658.

\bibitem{FMT89a}  F. M. Fern\'{a}ndez, Q. Ma, and R. H. Tipping, Tight upper
and lower bounds for energy eigenvalues of the Schr\"{o}dinger equation,
Phys. Rev. A 39 (1989) 1605-1609.

\bibitem{FMT89b}  F. M. Fern\'{a}ndez, Q. Ma, and R. H. Tipping, Eigenvalues
of the Schr\"{o}dinger equation via the Riccati-Pad\'{e} method, Phys. Rev.
A 40 (1989) 6149-6153.

\bibitem{F08}  F. M. Fern\'{a}ndez, Accurate calculation of eigenvalues and
eigenfunctions. I: Symmetric potentials, 2008. arXiv:0807.0655 [math-ph].

\bibitem{B75}  G. A. Baker Jr., Essentials of Pad\'{e} Approximants,
Academic Press, New York, San Francisco, London, 1975).

\bibitem{FT96}  F. M. Fern\'{a}ndez and R. H. Tipping, The Riccati-Pad\'{e}
quantization method for one-dimensional quantum-mechanical models, Can. J.
Phys. 74 (1996) 697-700.

\bibitem{F95}  F. M. Fern\'{a}ndez, Direct Calculation of Accurate Siegert
Eigenvalues, J. Phys. A 28 (1995) 4043-4051.
\end{thebibliography}
\end{document}